\documentclass[fleqn,usenatbib]{mnras}

\usepackage{newtxtext,newtxmath}

\usepackage[T1]{fontenc}

\DeclareRobustCommand{\VAN}[3]{#2}
\let\VANthebibliography\thebibliography
\def\thebibliography{\DeclareRobustCommand{\VAN}[3]{##3}\VANthebibliography}

\usepackage{graphicx}	
\usepackage{amsmath}	

\usepackage{rotating}
\usepackage{float}
\usepackage{subfigure}
\usepackage{graphicx,amsmath}

\usepackage{xcolor}
\usepackage{tikz}
\usetikzlibrary{shapes,arrows,backgrounds,calc,positioning}
\usetikzlibrary{calc,trees,positioning,arrows,chains,shapes.geometric,
decorations.pathreplacing,decorations.pathmorphing,shapes,
matrix,shapes.symbols}
\usetikzlibrary{calc}
\usepackage{array}
\usepackage{amsmath}
\usepackage{amsxtra}
\usepackage{amsfonts}
\usepackage{hyperref}
\usepackage{xspace}
\usepackage{multirow}
\let\new=\newcommand 
\new{\diff}{{\rm d}}

\title[Remnants of BH-BH mergers]{Where can we find the merger remnant in BH-BH mergers in Globular clusters?}

\author[Bhattacharyya \& Bagla]{
Dipanweeta~Bhattacharyya,$^{1,2}$\thanks{E-mail: dipanweeta.b@gmail.com}
Jasjeet~Singh~Bagla,$^{1}$\thanks{jasjeet@iisermohali.ac.in}
\\
$^{1}$Department of Physical Sciences, Indian Institute of Science Education and Research (IISER) Mohali, \\
Knowledge City, Sector 81, Sahibzada Ajit Singh Nagar, Punjab 140306, India\\
$^{2}$Department of Astronomy and Astrophysics, Tata Institute of Fundamental Research, \\ Homi Bhabha Road, Colaba
Mumbai 400005, India\\
}

\date{Accepted XXX. Received YYY; in original form ZZZ}

\pubyear{2023}

\begin{document}
\label{firstpage}
\pagerange{\pageref{firstpage}--\pageref{lastpage}}
\maketitle

\begin{abstract}
Mergers of black holes and other compact objects produce gravitational waves which carry a part of the energy, momentum, and angular momentum of the system. 
Due to asymmetry in the gravitational wave emission, a recoil kick velocity is imparted to the merger remnant. 
It has been conjectured that a significant fraction of the mergers detected so far reside in globular clusters.  
We explore the scenario where the merger remnant in a globular cluster is moving at a significant speed with respect to the binary that underwent merger.  
We study this in the situation when the kick velocity is higher than the escape velocity in the case of globular clusters assuming a Plummer density profile for the cluster. 
We study the evolution of the system to study the outcome: whether dynamical friction can trap the black hole within the globular cluster, whether the black hole escapes the globular cluster but ends up in the bulge, and lastly, whether the black halo becomes a halo object.  
We present results for an analysis based on orbital parameters of ten globular clusters using data from GAIA EDR3.
We find that if the kick velocity is smaller than $120$~km/s then a majority of remnant black holes end up in the bulge.
Note that our results in terms of where compact objects launched from a globular cluster end up are applicable to any mechanism, e.g., a compact object ejected due to three-body interactions.  
\end{abstract}




\section{Introduction}

The general theory of relativity predicts that every binary emits gravitational waves due to the second time derivative of the quadruple moment, and this leads to an eventual merger of the components of the binary. 
The first observational verification of the change in orbit due to the emission of gravitational waves was done by \citet{1989ApJ...345..434T}, while the first merger was observed in the gravitational waves (GW) window by \citet{2016PhRvL.116f1102A}, this was surmised to be a merger of two black holes (BH). 
Due to the anisotropic GW emission, a recoil kick is imparted to the merger product \citep{1983MNRAS.203.1049F, 2004ApJ...607L...5F, 2007PhRvL..98i1101G}. 
The estimated magnitude of these kick velocities ranges up to thousands of km/sec \citep{2007PhRvL..98w1102C, 2007PhRvL..98w1101G, 2019PhRvD.100j4039L, 2011PhRvL.107w1102L} and the magnitude depends on the parameters of the components of the binary.  
These kick velocities play a very important role in the evolution process of galaxies, star clusters as well as globular clusters \citep{2022PhRvL.128s1102V}. 
It also impacts BH formation and growth and GW astrophysics. 
Since the escape velocities of the Globular clusters (GCs) are fairly small  \citep{2021ApJ...914L..18D}, the merger remnant, when subject to a high kick velocity, can escape the cluster.
Hierarchical merging is a leading theory of the formation of massive BHs. 
When binary BHs (BBH) merge in dense clusters, the merger remnants can potentially merge with other BHs in the cluster, to form heavier BHs \citep{2021ApJ...918L..31M}. 
GCs are expected to be one of the prime sites for black hole merger events \citep{2019MNRAS.488.4370F, 2020ApJ...903...67M}.  
But here again, GW “kicks” might play an important role. 
It can cause some BHs to escape the cluster interrupting the process. 
\citet{2021ApJ...918L..31M} state that the BH kicks will not play the role of a major obstacle for hierarchical mergers if star clusters with escape velocities $\geq$ 200 km/s are abundant in the universe and BBHs with masses of that range in GWTC-2 mainly are formed in such clusters. 
However, typical globular clusters have much lower escape velocities and hence they are expected to retain only a small fraction of BH forming via mergers, as estimated by \citet{2021ApJ...918L..31M}.

According to \cite{2020PhRvL.124j1104V}, if precise measurements of BH parameters, mass, and spin are available, then using the fitting technique by numerical relativity \citep{2007ApJ...659L...5C}, it is possible to infer the kick velocity as a function of spin parameters and mass ratio of the BHs involved. 
\citet{2022PhRvL.128s1102V}, using numerical relativity surrogate models, put a constraint on the kick velocity for the event GW200129 and claim that there is $7.7\%$ probability for the black hole to be retained by globular or nuclear star clusters given this kick velocity. 
It is not an easy task to measure the magnitude of the kicks observationally. 
There are several ways to measure this - it can be measured by the Doppler shift in the gravitational wave of remnant BHs \citep{2016PhRvL.117a1101G, 2009JPhCS.154a2043F}, or by tracking certain frequency peaks in post-merger gravitational waveforms \citep{2020CmPhy...3..176C}, or by studying the higher order modes of the gravitational waveform which can infer the magnitude of these recoil kicks \citep{2018PhRvL.121s1102C}. 
But all of these needs very precise measurement to detect the kick magnitudes and are beyond the reach of detectors at present for typical events.

It is also important to figure out where the merged BH ends up if it gets ejected from GC and for that purpose, we need to track the trajectories of these ``kicked" BHs. 
This is important as many estimations suggest that some of the merger events seen by LIGO-VIRGO involve a later-generation black hole that has undergone several mergers to reach the present mass. 

In this paper, we describe a theoretical model for determining the fate of the BH kicked to a high velocity in a GC.  
This is a relevant approach to study the fate of a merger remnant (BH) of the BH-BH merger in GC. 
In \S 2, we describe our semi-analytic model to calculate the trajectory of these kicked BHs. 
We describe what are the possible sites where a BH ends up after getting kicked from the cluster.  
We show the results using different ranges of parameter sets and put the limits on the velocity for different cases in \S 3. We then use the data for 10 GCs \citep{2021RAA....21..173B, 2022arXiv221200739B}, to find the trajectories of the ejected BHs at different points in the orbit of the GC and calculate the fraction of the BHs which enter the bulge and when they escape bulge and end up being halo objects.  
In \S 4, we summarize and discuss the results. 

\section{Model}

We assume here that the process of dynamical friction reduces the kicked black hole's velocity after its formation in the merger.  
We assume that the merger happens at the center of the GC and what we call the kick velocity is the net velocity of the BH, i.e., the velocity of the center of mass of the merging system and the kick velocity added together. 
The analysis is independent of the specific processes that generate the kick.  
We solve the equation of motion of the kicked black hole to obtain the final velocity at the time when it exits the cluster and predict its final location. 
The analysis has been performed by assuming a Plummer density profile for GCs \citep{1911MNRAS..71..460P, 1997MNRAS.286..669G}.
Details of the model are given below. 

The density, $\rho_{P}$, potential $\Phi_{P}$ and the distribution function, $f_{P}$ for the Plummer model are:
\begin{equation}\displaystyle
\rho_{P}(r)=\frac{3M_{P}}{4\pi a_{P}^{3}}\bigg(1+\frac{r^{2}}{a_{P}^{2}}\bigg)^{-5/2},
\end{equation}
\begin{equation}\displaystyle
\Phi(P)=-\frac{GM_{P}}{\sqrt{r^{2}+a_{P}^{2}}},
\end{equation}
\begin{eqnarray}\displaystyle
f_{P}(\varepsilon) = \frac{24\sqrt{2}}{7\pi^{3}}\frac{a_{P}^{2}}{G^{5}M_{P}^{5}}\bigg(\frac{GM_{P}}{a_{P}}\bigg)^{-7/2}\varepsilon^{7/2},
\end{eqnarray}
where, the mass of the cluster is $M_{P}=2\pi \rho_{0} a_{P}^{3}$. 
$\rho_{0}$ is the central density of the cluster, and $a_{P}$ is the Plummer scale or core radius of the size of the cluster. 
$\varepsilon$ represents the energy in units of $\displaystyle \frac{GM_{P}}{a_{P}}$.

King profile is also another very commonly used profile for GCs, which is described by a length scale, mass (similar to the Plummer model), and a concentration parameter \citep{2008LNP...760..181K}. Here for simplicity, we are using the Plummer model for GCs since it has only two parameters describing the model.

For dynamical friction, we use the Chandrasekhar formula given by
\begin{equation}
\displaystyle \frac{\diff \vec{v}_{BH}}{\diff t} = -\frac{16 \pi^{2} G^{2} m_{*} (M_{BH}+m_{*}) \log \Lambda}{v_{m}^{3}}\int_{0}^{v_{BH}}f(v_{*}) v_{*}^{2} \diff v_{*} \vec{v}_{BH},
\end{equation}
where $M_{BH}$ is the mass under consideration, i.e, the mass of the kicked BH, $m_{*}$ is the mass of each star in the system, $v_{BH}$ is the velocity of the kicked BH, $\Lambda$ is the Coulomb logarithm and $f(v_{*})$ is the distribution function of the stars. 
$\Lambda$ has been calculated using $\displaystyle \frac{b_{max}}{b_{min}}$, where $b_{max}$ and $b_{min}$ are maximum and minimum impact factors for effective encounters. 
$b_{min}$ has been calculated using the formula $b_{min} = \displaystyle \frac{GM_{\bullet}}{v_{star}^{2}}$ and $b_{max}$ has been taken to be the size of the system. 
Then we solve this equation for $v_{BH}$, as a function of time using the parameters of the Plummer model. 

\begin{figure}
\Large
\begin{center}
\scalebox{0.35}{
\begin{tikzpicture}[node distance = 3cm, auto]
\tikzstyle{decision} = [diamond, draw,  
    text width=5em, text badly centered, node distance=3cm, inner sep=0pt]
\tikzstyle{block} = [rectangle, draw, 
    text width=10em, text centered, rounded corners, minimum height=3em]
\tikzstyle{line} = [draw, -latex']
\tikzstyle{cloud} = [draw, ellipse,node distance=3cm,
    minimum height=4em]
\node [cloud,fill=black] (para) {{\bf \color{white}\Huge Final location of the kicked Black Hole}};
\tikzstyle{block} = [rectangle, draw, 
    text width=10em, text centered, rounded corners, minimum height=3em]
\tikzstyle{line} = [draw, -latex']
\node [block,fill=cyan!20, below left of= para, node distance=7cm] (para3) {{\color{black}\Huge BH ends up crossing the Bulge}};
\tikzstyle{block} = [rectangle, draw, 
    text width=10em, text centered, rounded corners, minimum height=3em]
\tikzstyle{line} = [draw, -latex']
\node [block,fill=cyan!20, left of= para3, node distance=5cm] (para1) {{\color{black}\Huge BH remains in the Globular Cluster}};
\tikzstyle{block} = [rectangle, draw, 
    text width=15em, text centered, rounded corners, minimum height=3em]
\tikzstyle{line} = [draw, -latex']
\node [block,fill=cyan!20, right of= para3,node distance=6.5cm] (para6) {{\color{black} \Huge BH misses the bulge and becomes a halo object}};
\tikzstyle{block} = [rectangle, draw, 
    text width=15em, text centered, rounded corners, minimum height=3em]
\tikzstyle{line} = [draw, -latex']
\node [block,fill=cyan!20, right of= para6,node distance=7.5cm] (para4) {{\color{black} \Huge BH escapes the galaxy for very high kick velocity}};
\path [line, line width=0.8 mm] (para3)--node{}(para);
\path [line, line width=0.8 mm] (para6)--node{}(para);
\path [line, line width=0.8 mm] (para1)--node{}(para);
\path [line, line width=0.8 mm] (para4)--node{}(para);

\end{tikzpicture}

}
\end{center}
\caption{The four possibilities to locate the final position of the merged BH.}
\label{flowchart}
\end{figure}
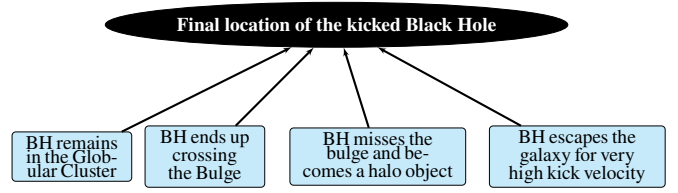

The star may slow down to remain trapped in the cluster, or it may escape from the cluster with a lower speed than the kick speed. 
If the BH escapes from the GC then the value of $v_{BH}$ at the time of ejection from the GC is denoted by $v_{ej}$.
Depending on the velocity of the BH after one crossing time, we have four possible scenarios shown in Fig. \ref{flowchart}.

If the velocity of the black hole comes down below the escape velocity, $v_{esc}$, within one crossing time, $t_{cross}$, then we assume that it stays within the GC itself. 
In this case, where the BH remains in the GC, we need to study and look for dynamical signatures in the GC: we can expect the BH to slow down due to continued dynamical friction and settle near the center of the cluster. 
The presence of an IMBH will affect the dynamics and the evolution of the GC. 

If the black hole velocity doesn't come down below $v_{esc}$ within one crossing time, then it escapes the cluster. 
Following \cite{2021ApJ...914L..18D}, we consider the escape velocity of the GCs to be $50$~km/sec, which is also within the specified range of central $v_{esc}$ for $141$ galactic GCs by \cite{2002ApJ...568L..23G}.
We consider a very simple scenario here: after getting ejected from the cluster, we assume that the BH moves in an elliptical orbit around the galactic center when the motion is bound which is again determined by the energy and we are assuming that the center of the bulge is at the focus of the ellipse.
We do not account for the mass distribution within the bulge in this approximation for computing the orbit of the GC or the BH.  

In the second case, BH escapes the GC but ends up crossing the bulge, in which case dynamical friction in the bulge is likely to trap it because the escape velocity for the bulge is much higher than that of the cluster.  
In case the kick velocities are intermediate between that required for escape from the GC and escape from the bulge, we expect a population of IMBH in the bulge resulting from this. 
We can expect signatures of microlensing \citep{2022arXiv220113296S, 2022arXiv220201903L} and also tidal interactions with the stars within the bulge. 

For the third case, BH misses the bulge and becomes a halo object. 
It can be expected to gradually slow down due to dynamical friction during disk crossings and get trapped in the disk and then sink towards the bulge. 
In this case also, signatures of microlensing are expected, though characteristics are expected to be different.

In the last and final case, the motion of the BH becomes unbounded and if the velocity is higher than the escape velocity of the galaxy it can escape the galaxy and move outside. 
In these cases, we can expect to find the BHs outside the galaxy in the inter-galactic medium.  This too can potentially be constrained by micro-lensing of cosmological sources like quasars, GRB afterglows, and high redshift supernovae. 
Thus observations of microlensing and dynamics of GCs may be used to constrain different scenarios and hence kick velocities.

\begin{figure}
    \centering
    \includegraphics[scale=0.4]{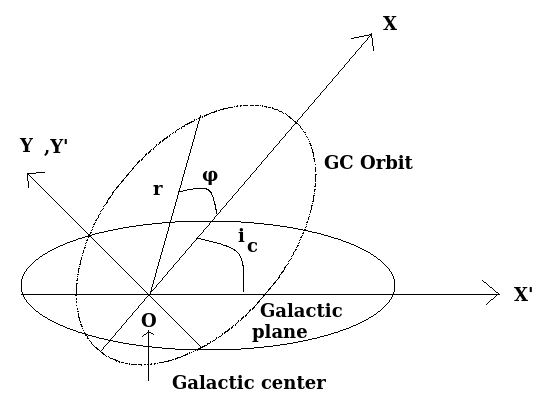}
    \caption{The orbit of GC, the galactic plane, and consideration of the X, Y, X$^{\prime}$ and Y$^{\prime}$ axes is shown along with the inclination angle, $i_{c}$.}
    \label{GC_Orbit}
\end{figure}

We use GAIA based estimates of the GC orbits around the bulge so that we can sample a realistic range of orbital parameters for GCs. 
We take the apoastron and periastron points of the GC orbit around the bulge, the inclination of their orbits, and also the time period of these orbits as the input parameters to calculate the position vector of these GCs at different points of their orbits. 
For simplicity, we assume the orbit to be elliptical. 
This assumption does not affect our results significantly.
To calculate the position of a GC in the orbit, we use the areal velocity equation:
\begin{equation}
\displaystyle \frac{1}{2}r^{2}\frac{\diff \phi}{\diff t} = \frac{\pi a b}{T},
\label{arealvel}
\end{equation}
where $a$ and $b$ are the semi-major and semi-minor axes of the elliptical orbit respectively, $\phi$ is the angular distance covered and $T$ is the total time period of the orbit. 
If the apoastron and periastron distances of the orbit are given by $d_{a}$ and $d_{p}$ respectively, then simplification of equation (\ref{arealvel}) gives
\begin{equation}
    \displaystyle \frac{\diff \phi}{\diff t} = \frac{\pi}{T r^{2}}(d_{a} + d_{p}) \sqrt{d_{a}d_{p}}.
\end{equation}
Then the angular momentum $L$ is calculated as
\begin{equation}
    L = m_{GC} r^{2}\frac{\diff \phi}{\diff t} = \frac{\pi m_{GC}}{T}(d_{a} + d_{p}) \sqrt{d_{a}d_{p}}
    \label{ang_mom_gc}
\end{equation}
The general equation of an ellipse is given by
\begin{equation}
    \displaystyle r = \frac{p}{1 + \varepsilon \cos \phi},
    \label{ellipse_eqn}
\end{equation}
where $\varepsilon$ is the eccentricity of the ellipse. 
$p$ is given by $\displaystyle \frac{L^{2}}{m_{GC} \times G M_{c}m_{GC}}$, where $M_{c}$ is the mass encircled by the orbit which we obtain by adding the halo and disk masses up to the distance $\displaystyle \frac{d_{a} + d_{p}}{2}$ to the bulge mass at the center of the galaxy (details are given below). 
Now using all these equations together and writing $\displaystyle \varepsilon = \frac{d_{a} - d_{p}}{d_{a} + d_{p}}$, we obtain an equation connecting $\phi$ and $t$, which provides the angular distance covered by the GC at any time in its orbit. 
\begin{equation}
    \displaystyle \int_{0}^{\phi} \frac{\pi^{3}}{G^{2}M_{c}^{2} T^{3}}\frac{(d_{a} + d_{p})^{3} (d_{a}d_{p})^{3/2}}{(1+\frac{d_{a} - d_{p}}{d_{a} + d_{p}} \cos \phi)^{2}} \diff \phi= \int_{0}^{t} \diff t
    \label{phi_t}
\end{equation}
After integrating equation (\ref{phi_t}) we can write $\phi(t)$, which in turn gives us $r(t)$ for the GC providing the position of the GC at any time in the orbit. 
Now to derive $\vec{r}$, we use the inclination $i_{c}$ of the orbit. 
If we consider the center of the bulge as the origin and consider the orbit to lie in the XY plane, then inclination $i_{c}$ provides the angle $\vec{r}$ makes with the galactic plane.
We note in passing that the most general description of the orbit will have more parameters though it is not obvious that these will add any new physical phenomena: we choose to keep the description simple (a closed ellipse with the given semi-major axis and eccentricity, and, the inclination to the galactic plane as parameters).  
We call the galactic frame as X$^{\prime}$Y$^{\prime}$ frame with the same origin and also assume that Y and Y$^{\prime}$ axes coincide. 
In XY frame, the components of $\vec{r}$ are $r\cos \phi$, $r\sin \phi$. 
The Z component is zero. 
Since the Y and Y$^{\prime}$ axes coincide, $\vec{r}$ in the X$^{\prime}$Y$^{\prime}$ frame can be considered to be rotated about the Y axis by an angle $i_{c}$. 
The orbit of GC, the galactic plane, and consideration of the axes are shown in Fig. \ref{GC_Orbit}. 

Therefore, $\vec{r}$ in X$^{\prime}$Y$^{\prime}$ frame is given as
\begin{equation}
    \vec{r} = r \cos \phi  \cos i_{c} ~\hat{i} + r \sin \phi ~ \hat{j} - r \cos \phi  \sin i_{c}  ~\hat{k}.
    \label{rvec}
\end{equation}
Differentiating equation (\ref{rvec}) with respect to time we find the velocity vector, $\vec{v}$ at position $\vec{r}$ of the GC. $\vec{v}$ in X$^{\prime}$Y$^{\prime}$ frame is found to be
\begin{eqnarray}
    \displaystyle \vec{v} = \bigg(-r(\phi) \cos i_{c} \sin \phi \frac{\diff \phi}{\diff t} + \frac{\diff r(\phi)}{\diff t} \cos i_{c} \cos \phi \bigg)~ \hat{i} \nonumber \\ + \bigg(r(\phi) \cos \phi \frac{\diff \phi}{\diff t} + \frac{\diff r(\phi)}{\diff t} \sin \phi \bigg)~ \hat{j} \nonumber \\  + \bigg(r(\phi) \sin i_{c} \sin \phi \frac{\diff \phi}{\diff t} - \frac{\diff r(\phi)}{\diff t} \sin i_{c} \cos \phi \bigg) ~\hat{k},
    \label{vvec}
\end{eqnarray}
where $\displaystyle \frac{\diff r(\phi)}{\diff t}$ is calculated as
\begin{equation}
    \displaystyle \frac{\diff r(\phi)}{\diff t} = \frac{p \varepsilon \sin \phi}{(1 + \varepsilon \cos \phi)^{2}} \frac{\diff \phi}{\diff t}.
    \label{drdt}
\end{equation}
To determine the mass encircled by the orbit, $M_{c}$, we use the mass densities of the disk and the halo given in \cite{2017MNRAS.465...76M}. 
The mass density of the disk, $\rho_{d}$, is given by
\begin{equation}
    \displaystyle \rho_{d} (R, z) = \frac{\Sigma_{0}}{2z_{d}}\exp \bigg(-\frac{|z|}{z_{d}} - \frac{R}{R_{d}}\bigg),
    \label{diskden}
\end{equation}
where for thin disk the values of scale height, $z_{d}$, central surface density, $\Sigma_{0}$ and scale length, $R_{d}$ are given by 300 pc, 896$M_{\odot}$pc$^{-2}$ and 2.5 kpc respectively. 
By performing volume integration of equation (\ref{diskden}) in cylindrical coordinate system for $z$ in the range -150 pc to +150 pc and $R$ in the range from 0 to $\displaystyle \frac{d_{a} + d_{p}}{2}$, we find the contribution of the disk mass in $M_{c}$. 
To find the halo mass contributing to $M_{c}$, we use the mass density of halo given by \cite{2017MNRAS.465...76M} as
\begin{equation}
    \displaystyle \rho_{h} = \frac{\rho_{0, h}}{x^{\gamma}(1 + x)^{3-\gamma}},
    \label{haloden}
\end{equation}
where $\displaystyle x = \frac{r}{r_{h}}$. 
The central mass density, $\rho_{0, h}$ and scale radius, $r_{h}$ are taken to be $0.00854$~M$_{\odot}$.pc$^{-3}$ and $19.6$~kpc respectively.  
$\gamma= 1$ for NFW profile \citep{1996ApJ...462..563N}, which is a very well-used dark matter profile in literature. 
Again by performing volume integration of equation (\ref{haloden}) in spherical polar coordinate by varying $r$ in the range 0 to $\displaystyle \frac{d_{a} + d_{p}}{2}$, we determine the contribution of halo mass in $M_{c}$.
These simplifying assumptions mean that our orbital reconstruction is not accurate.  
However, we expect that this has little impact on our estimations of the fraction that end up in the bulge or halo.

\begin{figure}
    \centering
    \includegraphics[scale=0.25]{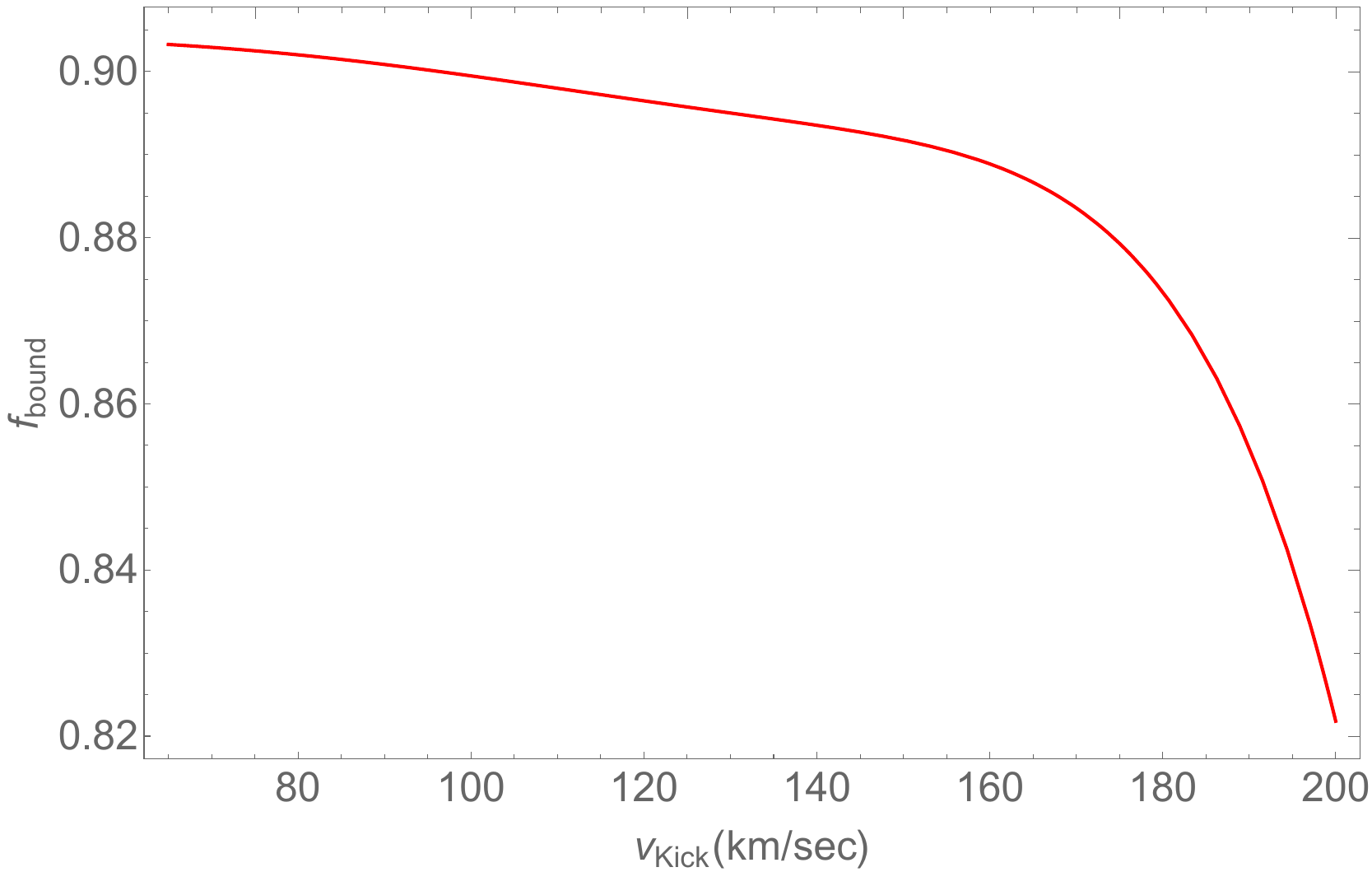}
    \caption{$f_{bound}$ as a function of $v_{Kick}$ $M_{BH} =  55 M_{\odot}$ calculated using orbital parameters of 10 Milky Way GCs listed in Table \ref{tableGC}.}
    \label{fbound}
\end{figure}

Now we focus on calculating the velocity and the trajectory of the BH that gets ejected from the GC after loosing some kinetic energy  due to dynamical friction in the GC. 
When calculated from the galacto-centric frame, the velocity of the ejected BH, $\vec{v}_{BHG}$ is given by $\vec{v} + \vec{v}_{ej}$, where $\vec{v}_{ej}$ is the velocity at which the BH gets ejected from the GC and $\vec{v}$ is the velocity of the GC calculated using equation (\ref{vvec}). 
This $\vec{v}_{ej}$ is a function of BH mass and the initial kick velocity, $v_{Kick}$ imparted to it. 
As there is no a priori constraint on the direction in which the BH is ejected from the cluster, we take it to be random. 
The initial position vector of the BH, $r_{BH}$, is the same as that of the GC. 
Then depending on its velocity and position, the motion of the BH will be bounded (it starts moving in an elliptical orbit around the galaxy center) or unbounded (it might escape the galaxy when the velocity exceeds the escape velocity of the galaxy). 
To check this, we calculate the total energy of the BH, $E_{BH}$ in the galactocentric frame. 
As long as $E_{BH}$ is negative, the motion will remain bounded. 
We find the solid angle within which the ejected BH is bound to the Galaxy.  
For a particular GC, depending on the mass and kick velocities, in some cases the motion can be unbounded. We show the average fraction of the bounded cases, $f_{bound}$ as a function of $v_{Kick}$ for 10 Milky Way (MW) GCs (listed in Table \ref{tableGC}; five GCs have $d_{p}$ less than 1 kpc, whereas the other five have $d_{p} >$ 1 kpc) in Fig. \ref{fbound}. 
We can see that BHs with higher $v_{Kick}$ are more probable to be unbounded after getting ejected from the GC and since we notice very less dependence on the mass of the BH, we show the result for only one mass value of BH ($M_{BH} =  55 M_{\odot}$). 
The average fraction of unbounded cases for this range of $v_{Kick}$ is around $\sim 10-18\%$. Here in this paper, we are focussing on the major fraction where the energy is negative implying a bounded motion around the galactic center.

Then we calculate the angular momentum, $L_{BH}$, of the BH by taking the cross product of $\vec{r}_{BH}$ and $\vec{v}_{BHG}$. 
For motion in an elliptical orbit under a central force, this is always conserved along with the energy.
For simplicity, we assume this to be the case here as any deviations from an elliptical orbit do not introduce changes in the conclusions. 
All these are used to calculate the eccentricity of the bound BH orbit, $\varepsilon_{BH}$, by using
\begin{equation}
    \displaystyle \varepsilon_{BH} = \sqrt{1+\frac{2E_{BH}L_{BH}^{2}}{M_{BH}(GM_{c}M_{BH})^{2}}}.
\end{equation}
Next, we estimate the distance of the closest approach in the BH orbit, $d_{BH} = a_{BH}(1-\varepsilon_{BH})$, where $\displaystyle a_{BH} = -\frac{GM_{c}M_{BH}}{2E_{BH}}$ is the semi-major axis of the BH orbit. 
If $d_{BH}$ comes out to be less than 1 kpc, then the ejected BH is expected to cross the bulge and if $d_{BH}$ is more than 1 kpc, we assume that it becomes a halo object.

Using GAIA EDR3 data of orbital parameters of globular clusters around the bulge \citep{2022arXiv221200739B}, we find the fraction of BHs which end up in the bulge, as also the fraction that avoid the bulge and become halo objects after being ejected from the GC by performing simulation over the whole range of the time period of the GC (so that it includes different positions of the GC on its orbit when the BH gets ejected from it) and for the BHs in the range of $\{5 M_{\odot},100 M_{\odot}\}$ and $v_{kick}$ in the range of $\{50,200\}$ km/sec. 
From our model analysis, we can also determine the angle between the velocity of the GC and $v_{ej}$, to understand at which angle it should get ejected to land up in the bulge or the halo. 
This can easily be calculated by taking a dot product of these two vectors. 

\begin{table*}
	\centering
	\caption{The list of 10 GCs used in our work with their orbital parameters taken from \citep{2022arXiv221200739B}.}
	\label{tableGC}
	\begin{tabular}{lcccr} 
		\hline
		GC & Apoastron ($d_{a}$) (kpc) & Periastron ($d_{p}$) (kpc) & Inclination ($i_{c}$) (deg) & Time period (T) (Myr)\\
		\hline
		NGC 6316 & 3.9 & 0.72 & 37 & 46 \\
		NGC 1851 & 19.9 & 0.17 & 97 & 228\\
		NGC 6440 & 1.5 & 0.05 & 104 & 14\\
        NGC 6453 & 2.6 & 0.2 & 84 & 32\\
        NGC 5286 & 13.0 & 0.54 & 116 & 144\\ \hline
        FSR 1758 & 12.0 & 3.31 & 148 & 152\\
        NGC 6426 & 16.7 & 3.28 & 26 & 204\\
        NGC 7078 & 10.9 & 3.77 & 29 & 144\\
        IC 4499 & 29.9 & 6.44 & 113 & 406\\
        NGC 5272 & 15.9 & 5.14 & 57 & 212\\
		\hline
	\end{tabular}
\end{table*}

In our simulation, we pick a random location of a given GC in its orbit.  
This is done by using the time spent by the GC at different locations as the distribution function: so the probability of the GC being located in any segment of the orbit depends on the time it is expected to spend in the segment as compared to the orbital period. 
The BH is then launched in a random direction with the kick velocity drawn from a distribution. 
The method described above is then followed to work out the fate of the BH.  
In order to estimate the fraction reliably, a large number of ejected black holes are considered in our simulations. 
This is then repeated for the sample of GCs.
The results of our simulations are presented in the following section. 

\section{Results}

\begin{figure}
\centering
\includegraphics[scale=0.25]{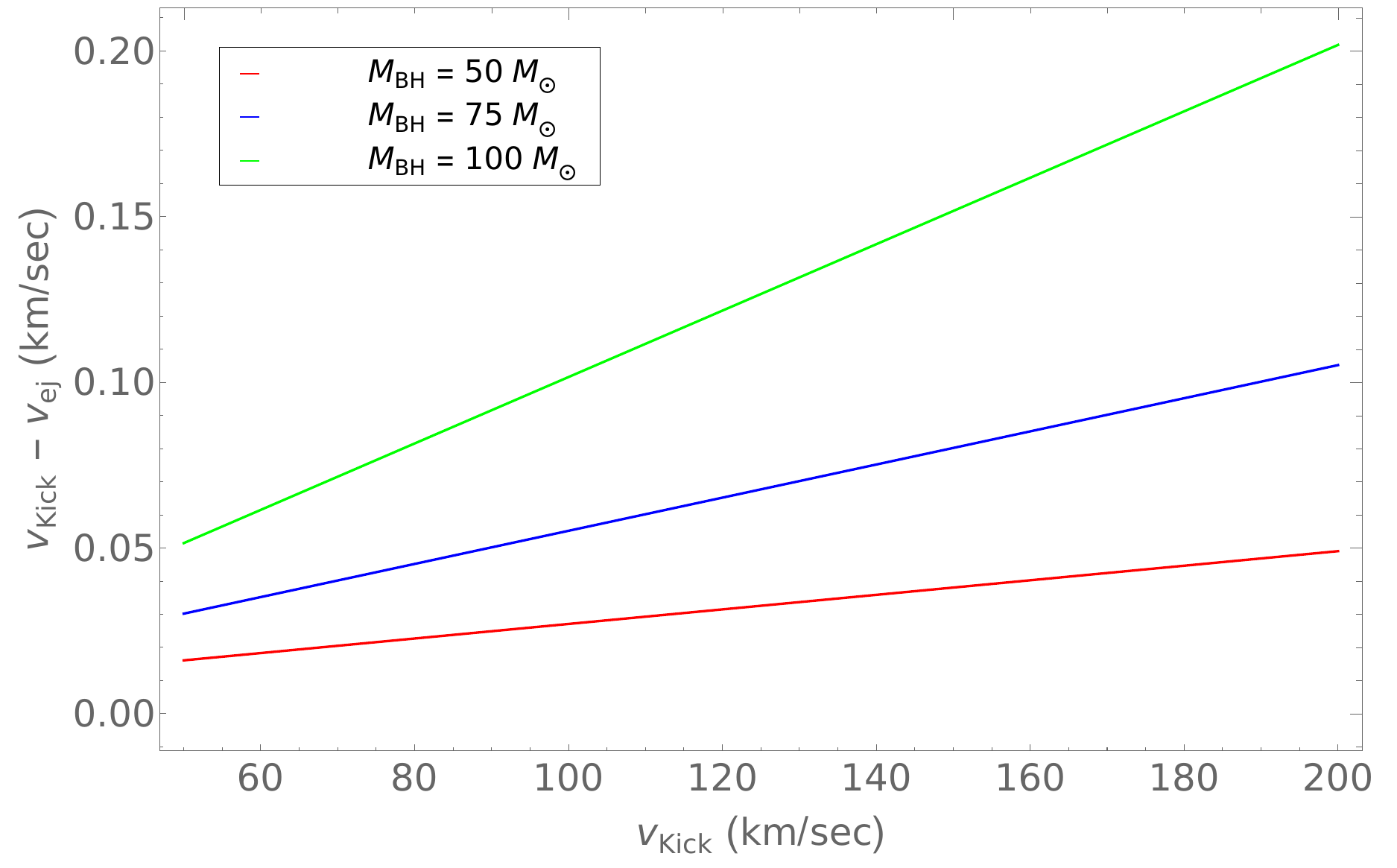}
\caption{The difference in $v_{Kick}$ and $v_{ej}$, the velocity at which the BH is ejected from the cluster after one $t_{cross}$ as a function of $v_{Kick}$ for different BH masses..}
\label{fig4}
\end{figure}

We show general cases in Fig. \ref{fig4} where we assume the BHs to get kick velocities of $50-200$~km/sec and we calculate the differences of $v_{Kick}$ and $v_{ej}$, which is the velocity after one crossing time, $t_{cross}$ for different BH masses.
We assume the core radii of the GC ($a_{P}$) to be 1 pc, mass of GCs to be $10^{5} M_{\odot}$, and radii of the GCs to be $\sim$ 10 pc.
We calculate the crossing time, $t_{cross}$, using the values of $v_{Kick}$ and and the size of the cluster and calculate the velocities of the BHs as these exit the cluster to check if those reduce below $v_{esc}$ or not.

Fig. \ref{fig4} shows that within one $t_{cross}$ the velocities of the BHs remain almost the same. 
The high-mass black holes tend to slow down more than the low-mass ones because of increased dynamical friction. 
But for all the cases, the decrease in velocity is too small to bring about a significant change in $v_{BH}$. 
Therefore, we conclude that if the BHs are imparted a $v_{Kick} > v_{esc}$, those will eventually escape the cluster, and the dynamical friction won't be able to stop the ejection.
As a result of the small radii of the GCs, the dynamical friction is not able to bind the BHs inside the clusters. This explains extremely low probability of finding the kicked BH (with $v_{Kick} > v_{esc}$) inside the GC. 
Thus, we conclude that only the merged BH which have $v_{Kick}<v_{esc}$ will only be found inside the cluster, whereas BH having $v_{Kick}>v_{esc}$ will  escape the cluster due to low $v_{esc}$ and small radius for clusters, which imply a short $t_{cross}$ for the BHs.

\subsection*{BH escapes from the cluster}

We here show the results for 10 MW GCs (listed in Table \ref{tableGC}; 5 GCs having $d_{p}$ less than 1 kpc, the assumed bulge radius, and 5 GCs having $d_{p}$ more than 1 kpc). 
For the first set of 5 GCs, at one point of time in their orbits, the BHs are ejected inside the bulge itself thereby increasing their probability to cross the bulge in their orbit. 
For the remaining 5 GCs, at every point in their orbits, the BHs are ejected outside the bulge radius decreasing their probability of having $d_{BH}$ within the bulge radius. 
But for some specific directions of the ejection, the BHs may have $d_{BH}$ within 1 kpc. 
We illustrate these points in Figs. (\ref{ngc6316}), (\ref{ic4499}) and (\ref{gc10}).

\subsection*{NGC 6316}

\begin{figure}
    \centering
    \subfigure[]{\includegraphics[scale=0.3]{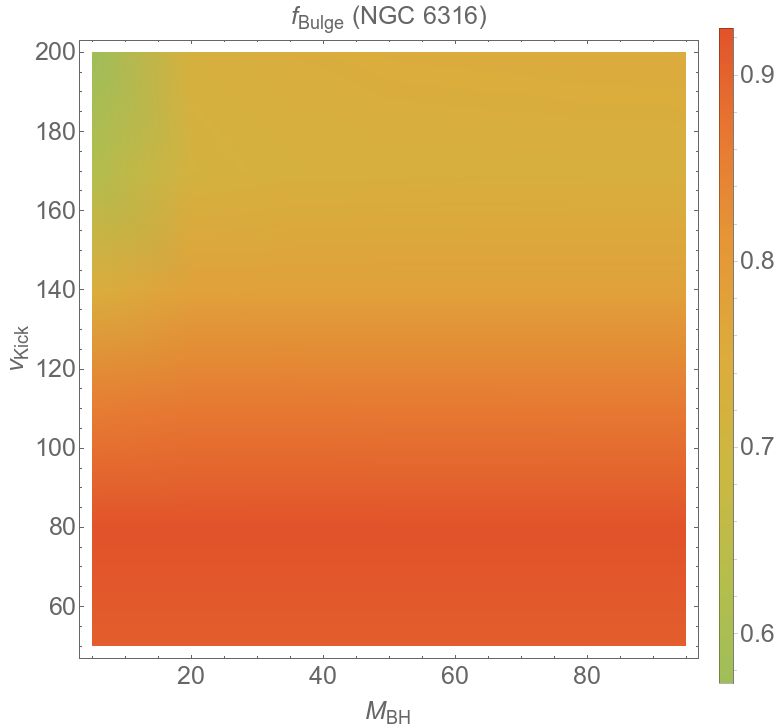}}\hspace{0.1cm}
     \subfigure[]{\includegraphics[scale=0.3]{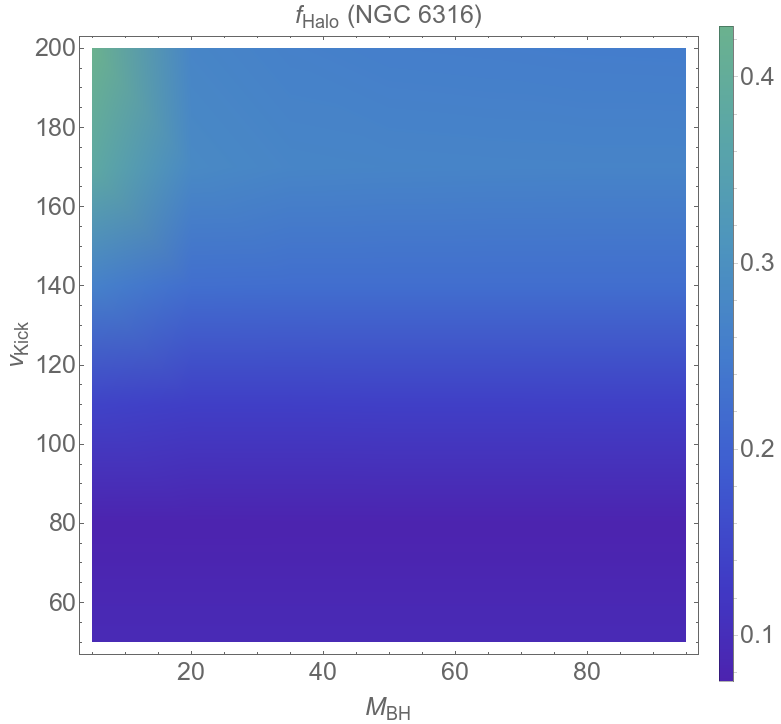}}
    \caption{The probable fractions of final positions of the ejected BHs from NGC 6316 in the bulge, $f_{bulge}$ (up) and (down) the halo, $f_{halo}$ for 0 to 46 Myr for $M_{BH}$ in the range $\{5 M_{\odot},100 M_{\odot}\}$ and $v_{kick}$ in the range of $\{50,200\}$ km/sec for the bounded cases.}
    \label{ngc6316}
\end{figure}

For NGC 6316, the mean apoastron and periastron distances are given as $3.9$~kpc and $0.72$~kpc with a time period of $46$~Myr. 
Therefore it belongs to the first set of GCs. 
The inclination angle is $37^\circ$. We obtain $d_{BH}$ from the formula given in \S 2 and generate the points for $0$ to $46$~Myr with BH masses in the range $\{5 M_{\odot}, 100 M_{\odot}\}$ and $v_{kick}$ in the range of $\{50,200\}$ km/sec for the bounded cases.  
We show contour plots of $f_{bulge}$ and $f_{halo}$ which indicate the probable fractions of final positions of the ejected BHs in the bulge and the halo respectively in Fig. \ref{ngc6316} for GC NGC 6316.

Fig. \ref{ngc6316}a shows that for a given $M_{BH}$, $f_{bulge}$ decreases with increasing $v_{Kick}$, which clearly indicates that for high-velocity BHs, the periastron of their orbits are bigger. Therefore they are less likely to enter the bulge. Also, the low-mass BHs with high velocities are more likely to enter the halo than the high-mass BHs with the same velocity.  This weak trend results from dynamical friction being more effective for higher mass BH.   
$d_{BH}$ also depends on the direction of the BH ejection and here we have generated the points considering all possible directions.
 We have generated a total of $\sim$ 2.75 $\times 10^{5}$ points to sample the solution space.
Also, in Fig. \ref{ngc6316}a, we notice less dependence on the BH mass. Fig. \ref{ngc6316}b shows the corresponding scenario for $f_{halo}$, which increases with increasing $v_{Kick}$.

\subsection*{IC 4499}

\begin{figure}
    \centering
    \subfigure[]{\includegraphics[scale=0.3]{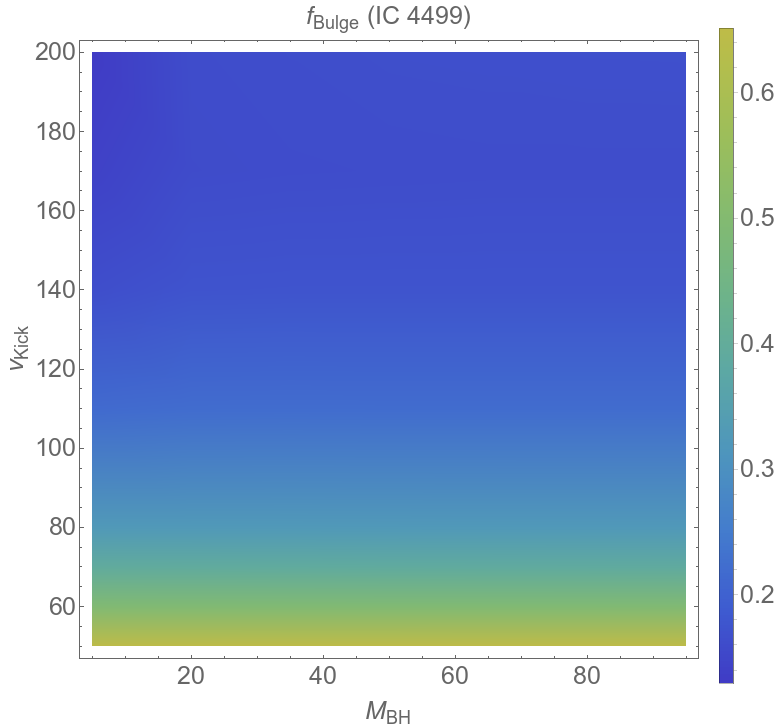}}\hspace{0.1cm}
     \subfigure[]{\includegraphics[scale=0.3]{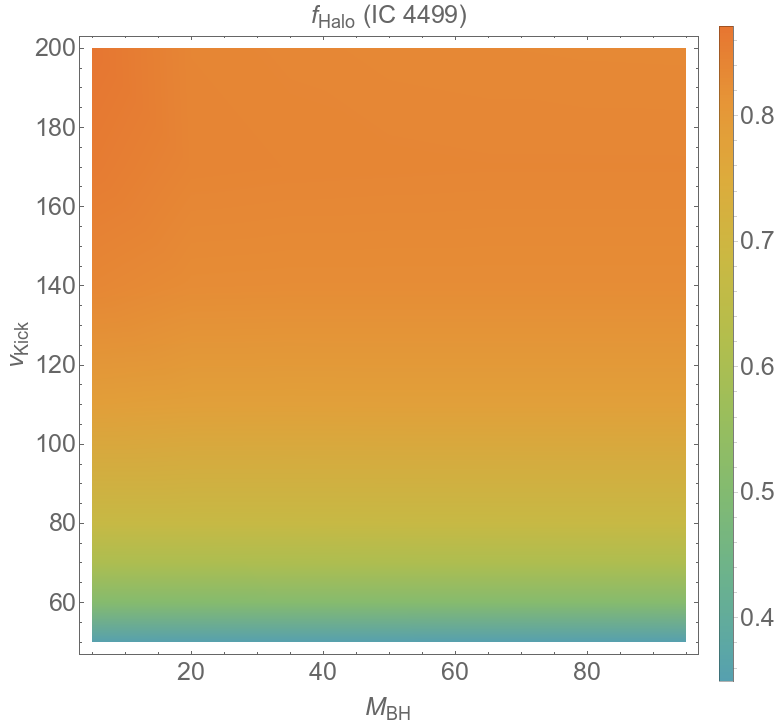}}
    \caption{The probable fractions of final positions of the ejected BHs from IC 4499 in the bulge, $f_{bulge}$ (up) and (down) the halo, $f_{halo}$ for 0 to 406 Myr for $M_{BH}$ in the range $\{5 M_{\odot},100 M_{\odot}\}$ and $v_{kick}$ in the range of $\{50,200\}$ km/sec for the bounded cases.}
    \label{ic4499}
\end{figure}

For IC 4499, the mean apoastron and periastron distances are given as $29,9$~kpc and $6.44$~kpc with a time period of $406$~Myr making it belong to the second set of GCs. The inclination angle is $113^\circ$.
We similarly  derive the $d_{BH}$ from the formula given in \S 2 and generate the points for $0$ to $406$~Myr with BH masses in the range $\{5 M_{\odot}, 100 M_{\odot}\}$ and $v_{kick}$ in the range of $\{50,200\}$ km/sec  for the bounded cases. 
We further show contour plots of $f_{bulge}$ and $f_{halo}$ which indicate the probable fractions of final positions of the ejected BHs in the bulge and the halo respectively in Fig. \ref{ic4499} for GC IC 4499.

In Figs. \ref{ic4499}a and \ref{ic4499}b, we can see a very similar trend to that of Fig. \ref{ngc6316}. 
The difference we see is that $f_{bulge}$ decreases ($f_{halo}$ increases) faster than NGC 6316. This happens due to the GC orbit which is larger than that of NGC 6316 and does not cross the bulge. 
Since for IC 4499, the BHs are ejected quite far from the galactic center, their probability of crossing the bulge is very small compared to NGC 6316, which orbits near the galactic center. 

\subsection*{Ten MW GCs}

\begin{figure}
    \centering
    \subfigure[]{\includegraphics[scale=0.3]{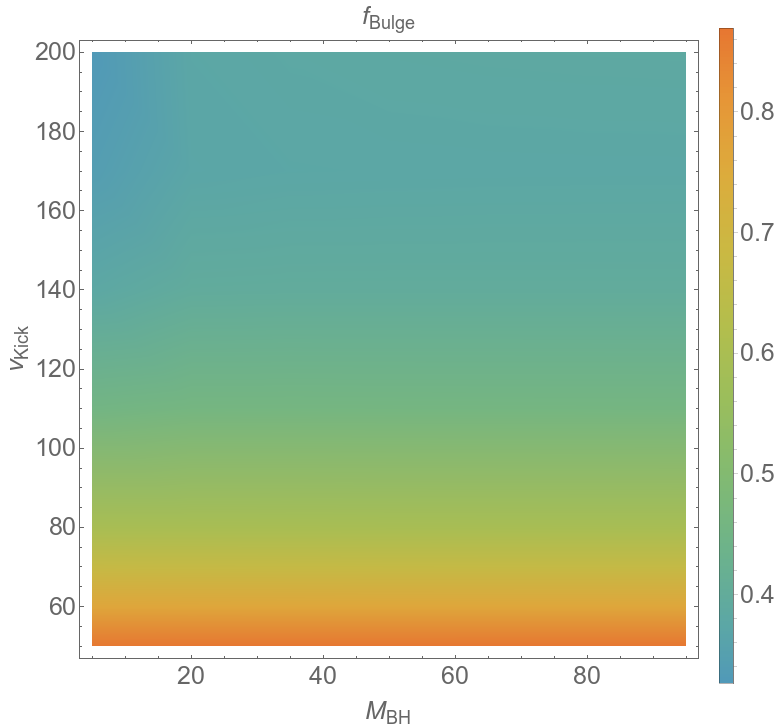}}\hspace{0.1cm}
     \subfigure[]{\includegraphics[scale=0.3]{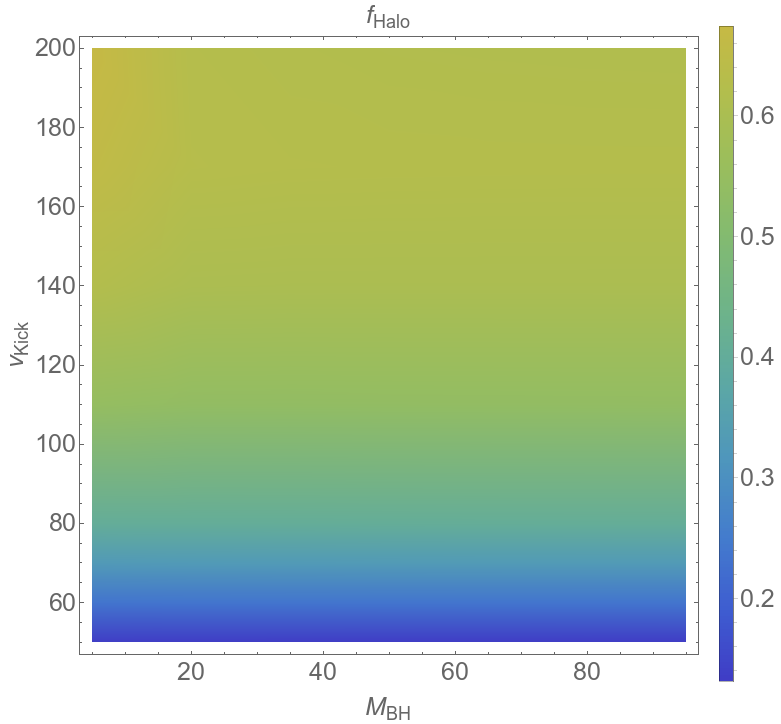}}
    \caption{The probable fractions of final positions of the ejected BHs from 10 GCs [listed in Table \ref{tableGC}] together, in the bulge, $f_{bulge}$ (up) and (down) the halo, $f_{halo}$ for their whole time periods for $M_{BH}$ in the range $\{5 M_{\odot},100 M_{\odot}\}$ and $v_{kick}$ in the range of $\{50,210\}$ km/sec for the bounded cases.}
    \label{gc10}
\end{figure}

Now we show the combined results for 10 GCs listed in Table \ref{tableGC}. 
These were chosen to represent the GC population in terms of their orbits, etc. 
We show the $f_{bulge}$ and $f_{halo}$ for all these 10 cases together to check which is the most probable destination of the ejected BHs  (for the bounded cases) for the majority of the cases in Fig. \ref{gc10}.

Figs. \ref{gc10}a and \ref{gc10}b show that upto $v_{Kick} \simeq$ 120 km/sec, the probability of BHs landing up in the bulge is high, whereas beyond this the probability of landing up in the halo is more. 
The whole analysis has been performed considering a fixed $v_{esc}$ for the GCs as $50$~km/sec. 
If for some cases, this value is higher $\sim 80 $~km/sec, then BHs having $v_{Kick} \sim 80$~km/sec, will be trapped in the GC itself. 
For such GCs a larger fraction of all escaping BH will end up in the halo. 

As mentioned above and in Fig. \ref{fbound}, the bulge and the halo fractions add up to around 90\% of all cases. 
In remaining cases the BH can be ejected from the galaxy. 
The unbound fraction increases with a larger kick velocity, as can be expected.

\section{Summary \label{s3}}

In this paper, we have explored the consequences of a massive compact object being kicked out of a globular cluster.  
This is a relevant study for merger of compact objects that have been observed using gravitational waves, if some of these events are taking place in globular clusters. 
Adding the fact that the anisotropic emission of gravitational waves leads to a kick imparted to the final black hole formed in the merger, we need to follow such systems to understand where these may end up. 
We model the system by assuming that globular clusters have a Plummer density profile.  
We consider dynamical friction and a simplified description of globular cluster orbits to estimate the fraction of black holes that remain within the globular cluster, the fraction that is likely to end up in the bulge, and the fraction that become halo objects. 
To do this, we consider orbits of ten globular clusters as representative of the entire population and use our semi-analytic model to estimate fractions as a function of the kick velocity and the black hole mass.
We summarize our key results below.
\begin{enumerate}
\item 
The velocity at which the black hole leaves a globular cluster is almost the same as the kick velocity, reduced by a very small fraction. 
This slowdown is due to dynamical friction and is more significant high-mass BBH merger remnant (Fig. \ref{fig4}).
\item 
If $v_{KicK}$ is beyond $v_{esc}$, the BHs are ejected from the GC and end up in the bulge or halo depending on its velocity, mass, and the point on the GC orbit at which it gets ejected. If $v_{KicK}$ is very high, in some cases, it might escape the galaxy as well. We show the cases where the motions of the BHs are bounded in Fig. \ref{fbound} and a fraction of the unbounded cases can leave the galaxy if the velocity is higher than the escape velocity of the galaxy.
\item 
We find that if the kick velocity is smaller than $120$~km/s then a majority of remnant black holes end up in the bulge (Fig. \ref{gc10}).
The tendency to end up in the bulge is pronounced for GCs whose orbit crosses the bulge (Fig. \ref{ngc6316}).
\item
For a higher kick velocity, the majority of black holes become halo objects (Fig. \ref{gc10}).
For much higher kick velocities, a significant fraction of remnant black holes escapes from the parent galaxy.
\end{enumerate}

These estimates have significant astrophysical implications.  
Using estimates of binary black hole merger rates, we expect a few hundred per Milky way type galaxy.  
Thus, depending upon the kick velocity, these could be in globular clusters, in the bulge, the halo, or even escape from galaxies and be intergalactic objects. 
Further, it has been estimated that the BH merger events make BH grow in a hierarchical manner and some of the observed events have resulted from BH formed as a result of a series of mergers.  
If this is the case then the kick velocity must be small in most cases if the mergers are happening in GCs over several generations. 

We find that a small fraction of BH is ejected from the galactic halo and this fraction increases with the kick velocity, being around $18\%$ for a kick velocity of $200$~km.s$^{-1}$.

A recent discovery of a candidate supermassive black hole streaming through the circumgalactic medium of a galaxy \citep{2023arXiv230204888V} indicates that large kick velocities are possible.  
Our analysis is agnostic as far as the origin of the kick is concerned.  
An object that is kicked out due to three-body or many body interactions will also move in exactly the same way.  
Thus objects thrown out due to three-body interactions can also escape the GC if their ejection velocity is more than the escape velocity.  

In situations where remnant BH are primarily halo objects or inter-galactic objects, microlensing of background sources is the primary observational constraint.  
A detailed dynamical study is required for scenarios where the BH remains in the globular cluster or becomes a halo object. 
The effect of the remnant BH on the dynamics of the globular cluster can also be a useful probe.  
Some of these questions are the subject of our ongoing work and will be reported elsewhere. 

Observational constraints from micro-lensing on compact bulge/halo/intergalactic objects will be of immense significance in constraining the scenarios. 

The impact of fast moving BH on the dynamical state of GCs also requires a detailed study as we may be able to see signatures of past events.

\section*{Acknowledgements}

We would like to thank Professor Kulinder Pal Singh for his helpful suggestions. 
This work has made use of NASA's Astrophysics Data System Bibliographic Service.

\section*{Data Availability}

The original data generated in simulations in this paper will be made available on request to the corresponding author.



\bibliographystyle{mnras}
\bibliography{references} 








\bsp	

\label{lastpage}
\end{document}